# Growth of Large-area Single- and Bi-layer Graphene by Controlled Carbon Precipitation on Polycrystalline Ni Surfaces

*Alfonso Reina[1], Stefan Thiele[2], Xiaoting Jia[1], Sreekar Bhaviripudi[3], Mildred S. Dresselhaus[3,4], Juergen A. Schaefer[2], and Jing Kong[3,\*]*

[1]Department of Materials Science and Engineering, Massachusetts Institute of Technology, Cambridge, Massachusetts 02139, USA

[2]Institut für Physik and Institut für Mikro- und Nanotechnologien, Technische Universität Ilmenau, Ilmenau, Germany 98684

[3]Department of Electrical Engineering and Computer Sciences, Massachusetts Institute of Technology, Cambridge, Massachusetts 02139, USA

[4]Department of Physics, Massachusetts Institute of Technology, Cambridge, Massachusetts 02139, USA

Address correspondence to jingkong@mit.edu

ABSTRACT

We report graphene films composed mostly of one or two layers of graphene grown by controlled carbon precipitation on the surface of polycrystalline Ni thin films during atmospheric chemical vapor deposition (CVD). Controlling both the methane concentration during CVD and the substrate cooling rate during graphene growth can significantly improve the thickness uniformity. As a result, one- or two- layer graphene regions occupy up to 87% of the film area. Single layer coverage accounts for 5–11% of the overall film. These regions expand across multiple grain boundaries of the underlying polycrystalline Ni film. The number density of sites with multilayer graphene/graphite (> 2 layers) is reduced as the cooling rate decreases. These films can also be transferred to other substrates and their sizes are only limited by the sizes of the Ni film and the CVD chamber. Here, we demonstrate the formation of films as large as 1 in$^2$. These findings represent an important step towards the fabrication of large-scale high-quality graphene samples.

Exploring ways to synthesize graphene which allow scalability, have low fabrication cost and facilitate integration with other materials, could play an important role in both fundamental research and the realization of future graphene applications. Several graphene synthesis approaches have been developed, including: (1) exfoliation methods (both mechanical [1, 2] and chemical [3-8]); (2) graphitization of SiC surfaces [9, 10]; (3) graphene precipitation/deposition on transition metal surfaces [11, 12]; and (4) gas phase/substrate-free formation of graphene sheets [13]. Procedures such as (1) and (4) produce free-standing graphene isolated from any substrate which enables the integration of graphene with other materials. Methods like (2) and (3) produce graphene bound to a specific substrate which limits the flexibility of these approaches. Recently it has been shown that graphene films can be grown by ambient pressure CVD on thin films of transition metals and isolated from their growth substrate [14-17]. This approach is promising for generating large scale graphene on a wide range of substrates. However, these films vary in thickness from 1 to ~10 layers across their area [14-16]. Here, we present an important advance to further improve the thickness uniformity of these films. We show that the area of multilayer graphene regions on the film can be reduced and the regions with single- or bi-layer graphene (1–2 LG) can be increased to 87% of the overall film area.

The precipitation of monolayer and multilayer graphene from bulk transition metals is widely known [18, 19]. It occurs due to the temperature-dependent solubility of carbon in transition metals. This concept has recently been used to produce 1–2 LG under either vacuum conditions [11, 20] or in an ambient pressure CVD process [14-17]. In our process, carbon is introduced into the bulk of thin (~500 nm) Ni films by decomposing a highly diluted hydrocarbon gas ($CH_4$). Graphene precipitation is promoted on the surface of the Ni films upon cooling of the Ni–C solid solution. A summary of the three-stage process that is utilized is shown in Fig. 1. The samples are heated to 900 ˚C and annealed for 20 minutes at this temperature (stage 1) under Ar and $H_2$ in order to smooth the Ni surface and to activate Ni grain growth. During stage 2, $CH_4$ (typically around 0.5–1 vol. % by controlling the flow rate) along with $H_2$ is allowed to flow over the Ni surface at 1000 ˚C. $CH_4$ begins to decompose catalytically on the surface of Ni [21, 22] and carbon diffuses into the Ni film. After 5 minutes of $CH_4$ exposure, the Ni film is cooled down (stage 3) under Ar, $H_2$ and the same $CH_4$ concentration (see Fig. 1 and Table 1 for exact flow rates). Based on previous models of non-equilibrium carbon segregation in Ni [19, 23], when the Ni–C solution is cooled down, graphene precipitates on the surface of the Ni film. We report that by controlling the amount of methane during our process and reducing the rate of substrate cooling in stage 3, it is possible to obtain graphene films consisting of mostly 1–2 LG (see Fig. 2). The Ni films utilized are polycrystalline due to their deposition method (E-beam evaporation or sputtering) with a thickness of ~500 nm. The role of the Ni grain size on the thickness variation of the graphene films has been discussed previously [24]. In this work, Ni films were deposited under conditions which give two different average Ni grain sizes of a few microns or a few tens of microns after annealing (i.e., after stage 1). Results are compared using Ni substrates with both grain sizes. The use of these types of Ni films is attractive due to their relatively simple fabrication and low cost compared to single crystalline Ni. Transfer of the resulting graphene was done by wet-etching of the Ni film with a 3 wt. % aqueous solution of hydrochloric acid. Before etching, a layer of poly(methyl methacrylate) (PMMA) was spin-coated on the surface of the graphene film to serve as a support. The resulting PMMA/graphene layer was then manually laid on the target substrate ($SiO_2$/Si or TEM grids). The PMMA was finally removed by exposure to acetone in liquid or vapor form.

Two types of graphene films (A and B, shown in Fig. 2) with contrasting thickness variations can be obtained by controlling the methane concentration during CVD ($X_{methane}$) and the rate of cooling (*dT/dt*). Table 1 shows the conditions under which each type of film can be grown. Films of type A consist mostly of multilayer graphene and are grown with high $X_{methane}$ (0.7 vol. %). It is observed that the Ni grain size plays a critical role in the thickness variation of the film, as also reported previously [23]. Multilayer graphene with more than two graphene layers (2+ LG) tends to segregate around the grain boundaries of the polycrystalline Ni film (Figs. 2a and 2b). The thinnest regions that were identified (1–2 LG) grow at the center of the large Ni grains of the catalytic Ni film. These observations suggest that Ni grain boundaries act as preferential nucleation sites for multilayer graphene or graphite. This can be explained by the fact that impurities in transition metals tend to segregate at grain boundaries [25, 26]. On closer scrutiny, comparison of Fig. 2a and Fig. 2b reveals that multilayer graphene was present at almost all the Ni grain boundaries, suggesting that the density of nucleation sites for graphene precipitation is high (as compared to graphene film type B which is discussed later). For a $CH_4$ concentration of 0.7 %— which always results in the growth of type A films—the size of 1–2 LG is independent of the cooling rate (see the summary in Table 2), but does depend strongly on the average grain size of the Ni film used for the synthesis. Therefore, Ni films with different average grain sizes produce 1–2 LG regions of different sizes [24].

Graphene films with their area consisting mostly of 1–2 LG (type B in Figure 2) are grown by using intermediate $X_{methane}$ values (0.5–0.6%) and low cooling rates (*dT/dt* <25 °C / min). For this case, the film thickness variation obtained is significantly different from that obtained with higher $X_{methane}$ (0.7 %). It is observed that not all grain boundaries on the polycrystalline Ni show the nucleation of multilayers (Figs. 2c and 2d), resulting in an increase of the area fraction covered by 1–2 LG. AFM, TEM, Raman spectroscopy and optical microscopy were used to characterize these films (Figs. 2e–2h). The heights of 1- and 2-LG on $SiO_2$/Si as measured by AFM are 0.72 and 1.16 nm, respectively (Fig. 2e). TEM confirmed that most of the film area consists of 1–2 LG (Figs. 2f and 2g). The Raman G´ band (~2700 cm$^{-1}$) of 1- and 2-LG always has a single Lorentzian lineshape. For both cases, the linewidth usually lies in the range 30–40 cm$^{-1}$, suggesting the absence of interlayer coupling. Furthermore, the relative intensity of the G´ and G bands varies randomly between 1- and 2-LG regions, possibly due to differences in doping levels [27]. Also, no difference in G´ frequency is observed between the 1- and 2-LG regions probed [28]. Therefore, it is not possible to distinguish between 1- and 2-LG using Raman spectroscopy alone. This is better done by optical microscopy (see discussion below and Electronic Supplementary Material (ESM)) or direct observation in a TEM. Four point probe measurements of the sheet resistance of the films yield values of ~0.5–1 kΩ / sq and 3–5 kΩ / sq for films of type A and B, respectively. The difference in sheet resistance is attributed to the conduction through multilayer graphene which should have a larger contribution to the film conductivity in the case of films of type A.

The differences in the number of multilayer graphene sites between films of types A and B can be explained in terms of the differences in the methane concentrations and cooling rates used. Lower methane concentrations will result in relatively low carbon concentrations in the Ni film. Consequently, this will promote a reduction of carbon segregation on the Ni surface during the cooling stage. The amount of segregation, for a given change in temperature $dT/dt$, depends on the magnitude of the solute over-saturation (which should be directly related to the methane flow rate) [29-31]. On the other hand, decreasing $dT/dt$ may promote segregation under conditions closer to equilibrium, therefore reducing the density of multilayer sites [32]. Note that with 0.5% methane concentration, only films of type B were obtained, whereas if the methane concentration was increased to 0.7%, only films of type A were obtained. This is consistent with our ideas discussed above. Table 1 shows that at 0.6% methane, there was a transition from A to B type of film growth as the cooling rate was decreased. However, it was found that this methane concentration resulted in a partial graphene coverage of the Ni surface if high cooling rates (33–100 ˚C / min) were used (see Table 2). For the same methane concentration, using low cooling rates (<25 ˚C / min) resulted in full coverage but with an inhomogeneous density of multilayer sites. The best control over both graphene coverage and homogeneous density of multilayer sites was accomplished with methane concentrations of 0.5 and 0.7% for films of type A and B, respectively (see Table 2).

It was found that in type B films, grown with $X_{methane}$ = 0.5%, the area covered by 1–2 LG ($\theta_{1-2LG}$) was dependent on the cooling rate (Table 2). Figure 3 shows that decreasing the cooling rate below 25 °C / min during the segregation step further increased $\theta_{1-2LG}$ in type B films ($X_{methane}$ ~0.5%). In addition, the slower the cooling rate, the fewer the number of nucleation sites of multilayer graphene ($\rho_{2+LG}$). The decrease in the density of multilayer sites can be also attributed to a reduction of the segregation rate caused by the lowering of $dT/dt$. At low segregation rates, carbon atoms can diffuse for longer times before they coalesce to form graphene (diffusion limited nucleation) [30]. It can also be observed that as $dT/dt$ decreased, not only did $\rho_{2+LG}$ decrease but the thickness of the multilayer pieces increased. This can be seen by the increase in the number of yellow or white regions (graphite) and the reduction in the number of purple or blue regions on the graphene film (Figs. 3a–3c). This suggests that $dT/dt$ may only have an effect on the density of multilayer sites, and not on the amount of carbon segregating. Consequently, in the case of our slowest cooling rates, if fewer nucleation sites are available for the same amount of carbon segregating at the surface, an increase in the thickness of the multilayer graphene regions must be expected.

The cooling rate of the Ni film during graphene precipitation was used to obtain films with up to 87% of their area ($\theta_{1-2LG}$ = 0.87) composed of no more than two layers of graphene (of which the single layer area made up 5–11% of the total film area). The area fraction $\theta_{1-2LG}$ increased as the cooling rate was decreased and it can be tuned from 0.60–0.87 for CVD processes using $X_{methane}$ ~0.5% (Fig. 4). The density of sites consisting of multilayer graphene with more than two layers, $\rho_{2+LG,}$ can be decreased by 50% on going from the highest to the lowest cooling rate tested (Fig. 4). The quantification of the area percentage plotted in Fig. 4 was done by comparing optical images of the graphene films on SiO$_2$/Si with bare SiO$_2$/Si substrates. Each pixel of the optical images can be expressed in the RGB (Red Green Blue) color model [33] which is used for image display and representation in electronic systems. In this model, the color of each pixel in an image is represented by the intensities of the three primary colors— red, green and blue, hence its name. When graphene is present on 300 nm $SiO_2$/Si, it creates an enhanced absorption at wavelengths around 500 nm [34, 35] corresponding to the color green. Therefore, the green component, {G} of our optical images can be used to identify the contrast created by the CVD graphene film with respect to the underlying $SiO_2$ (Fig. 4a). This enables us to identify regions with down to 1- and 2-LG in an automated way (see ESM). Such a contrast in {G} was also measured for pieces of exfoliated graphene (from highly oriented pyrolytic graphite (HOPG)) on $SiO_2$/Si and was used as a calibration for the identification of 1- or 2- LG derived from our CVD process (see ESM). The coverage, $\theta_{1-2LG}$, plotted in Fig. 4b represents the fraction of pixels identified as containing no more than two graphene layers (pink background in images of Fig. 3). Optical images at 50x magnification, with 3900 by 3090 pixels (289 by 229 $\mu m^2$), were used for this analysis. It was also observed that $\theta_{1-2LG}$ is independent of the grain size of the Ni film used to synthesize graphene (Figs. 4b and 4c). Two Ni grain sizes ($L_1$ and $L_2$) were used in our experiments and their images are shown in Figs. 4d and 4e. Optical images of graphene grown on both Ni grain structures and transferred to $SiO_2$/Si are shown in Figs. 4f–4i. This comparison is important since the grain sizes of transition metal thin films vary depending on film thickness, residual stress and deposition conditions [36, 37]. Lastly, these films are also transferable to other substrate materials, similar to the way such transfers have been reported previously [15]. Graphene films of up to 1 $in^2$ in size and with high area fractions of 1–2 LG have been fabricated (Fig. 4j). Their sizes are limited only by the size of the Ni film used and the CVD chamber size.

In conclusion, we have demonstrated the possibility of growing graphene films with up to 87% of their area composed of no more than two graphene layers and which can also be transferred to insulating substrates. This was accomplished by controlling both the carbon concentration and the substrate cooling rate during the CVD process. Under a suitable carbon concentration (0.5% $CH_4$ in our case), the cooling rate can be utilized to decrease the number of nucleation sites of multilayer graphene on the film (by a factor of two) and to increase significantly the area covered by sections with 1–2 LG. Further quantitative analysis (for example, the carbon concentration inside the Ni film for substrates treated with different $CH_4$ exposures and cooling rates) is currently being undertaken in order to gain a deeper understanding of this process. Nevertheless, our results suggest the possibility of dramatically improving the thickness uniformity of graphene films by controlling the process parameters in our method. Therefore, ambient pressure CVD may be a viable route to control the growth of single graphene layers over large scales.

ACKNOWLEDGMENT

This work was partly supported by the Materials, Structures and Devices (MSD) Focus Center, one of the five centers of the Focus Center Research Program, a Semiconductor Research Corporation program. Support from NSF/CTS 05-06830 (X. J. and M. S. D) and NSF/DMR07-04197 (A. R. and M. S. D) is also acknowledged. Raman measurements were carried out in the George R. Harrison Spectroscopy Laboratory supported by NSF-CHE 0111370 and NIH-RR02594 grants. The authors acknowledge Mario Hofmann for help in preparation of graphic illustrations for potential cover art and GerardoMartinez and Arturo Ponce Pedraza (Laboratory of Microscopy at the Research Center for Applied Chemistry (CIQA). Saltillo, Coahuila, Mexico.) for providing images for the potential cover art.

**Electronic Supplementary Material:** Details of the automated recognition of regions of one and two graphene layers by computer programs using optical images are available in the online version of this article at www.thenanoresearch.com and are accessible free of charge.

TABLES

TABLE 1. Types of graphene films obtained at different methane concentrations and cooling rates.

| $X_{methane}$ (vol. %) | $dT/dt$ (°C / min) | | | | | | |
|---|---|---|---|---|---|---|---|
| | 100.0 | 33.0 | 25.0 | 16.6 | 8.3 | 5.5 | 4.2 |
| 0.4 | No graphene film | | | | | | |
| 0.5 | No graphene film | | B | | | | |
| 0.6 | A | | B | | | | |
| 0.7 | A | | | | | | |

A=films of type A (Figs. 2a and 2b). B=films of type B (Figs. 2c and 2d)

TABLE 2. Description of films obtained with different $CH_4$ concentrations and cooling rates.

| $X_{methane}$ (vol. %) | Regime of $dT/dt$ (°C / min) | |
| --- | --- | --- |
| | High (100.00 °C / min) | Low (<25 °C / min) |
| 0.5 | No graphene film | **B**<br><br>$\theta_{1\text{-}2LG}$ depends on <u>cooling rate</u><br><br>$\rho_{2+LG}$ is <u>homogeneous across Ni surface</u><br><br><u>Full</u> coverage of the graphene film on the Ni surface |
| 0.6 | **A**<br><br>$\theta_{1\text{-}2LG}$ similar to the <u>size of Ni grains</u><br><br>$\rho_{2+LG}$ is <u>homogeneous</u> across the graphene film<br><br><u>Partial</u> coverage of the graphene film on the Ni surface | **B**<br><br>$\theta_{1\text{-}2LG}$ depends on <u>cooling rate</u><br><br>$\rho_{2+LG}$ is <u>inhomogeneous</u> across the graphene film<br><br><u>Full</u> coverage of the graphene film on the Ni surface |
| 0.7 | **A**<br><br>$\theta_{1\text{-}2LG}$ similar to <u>the size of Ni grains</u><br><br>$\rho_{2+LG}$ is <u>homogeneous</u> across the graphene film<br><br><u>Full</u> coverage of the graphene film on the Ni surface | **A**<br><br>$\theta_{1\text{-}2LG}$ similar to <u>the size of Ni grains</u><br><br>$\rho_{2+LG}$ is <u>homogeneous</u> across the graphene film<br><br><u>Full</u> coverage of the graphene film on the Ni surface |

A=films of type A (Figs. 2a and 2b). B=films of type B (Figs. 2c and 2d). $\theta_{1\text{-}2LG}$ = area fraction occupied by one or two graphene layers. $\rho_{2+LG}$= number density of multilayer sites with more than two graphene layers.

FIGURE CAPTIONS

**Figure 1** Illustration of the graphene growth process and its different stages (1–3). 1. The Ni film deposited on $SiO_2$/Si is heated to 900 ˚C and annealed for 20 minutes under flowing $H_2$ and Ar (400 and 600 standard cubic centimeter per minute (sccm), respectively). Here, Ni grain growth and surface smoothening occurs. 2. Exposure to $H_2$ and $CH_4$ for 5 minutes. The flow rate of $H_2$ is always 1400 sccm in every run. The flow rates of $CH_4$ used in the results presented in Table 1 are 6, 7, 8 and 10 sccm corresponding to concentrations of 0.4, 0.5, 0.6 and 0.7 vol. %, respectively. $CH_4$ is decomposed catalytically and the carbon produced is incorporated into the Ni film. 3. The substrate is cooled down from 1000 ˚C to 500 ˚C under Ar, $H_2$ (700 sccm for both gases) and the same flow of $CH_4$ is used as in stage 2. Times for this step are 15 to 120 minutes corresponding to cooling rates between 33 and 4.2 ˚C / min. At 500 ˚C, the sample is taken out of the tube furnace and cooled rapidly to room temperature. For the case of a cooling rate of 100 ˚C / min, the sample is simply taken out of the furnace and cooled down to room temperature.

**Figure 2** Two types of graphene films (types A and B) and their characterization. (a, b) Type A film with low coverage of one to two layer regions (low $\theta_{1\text{-}2LG}$). (c, d) Type B film with high coverage of one to two layer regions. (a) and (c) are optical images of the graphene films on Ni, (b) and (d) are optical images of the graphene films transferred to $SiO_2$/Si. Transfer to $SiO_2$/Si enables thickness analysis by optical contrast. (e) AFM image of a 1–2 layer region on $SiO_2$/Si of a type B film. Inset shows the cross sectional height of 1 and 2 LG measured along the lines shown in the AFM topographical image. (f) SEM image of a 1–2 LG region of a type B film transferred to a TEM grid for thickness analysis. The regions remain freestanding across the circular openings of the grid. Dark areas suggest that the film broke at those sites. (g) TEM image of a region consisting of 1–2 LG in a type B film (pink background in (d)). (h) Representative Raman spectra collected from a type B film at regions consisting of 1–2 LG (shown in red) and 2+ LG (~5L, shown in blue). The G´ peak at ~2700 $cm^{-1}$ for 1–2 LG layers is a single Lorentzian peak. The Raman spectra of graphite pieces found in the film

(yellow clusters in optical image (d)) are shown in green. The laser wavelength used was 514 nm with a laser power of 1 mW and acquisition time of 5 s .

**Figure 3** Effect of the cooling rate on type B films which were grown with a $CH_4$ concentration of 0.5%. (a–c) Optical images of graphene films transferred to $SiO_2$/Si grown with decreasing cooling rate as indicated. The number of nucleation sites with multilayer graphene decreases as the cooling rate decreases, leading to an increase of the 1–2 LG region (pink background). Scale bars are all 25 µm.

**Figure 4** Quantification of single and bilayer graphene coverage of graphene films grown on Ni with different grain sizes ($L_1$ and $L_2$). (a) Optical recognition of 1- and 2- LG with the {G} values extracted from the RGB image of graphene films on $SiO_2$/Si. {G} decreases in a stepwise manner from bare $SiO_2$ to one and to two graphene layers (inset in (a)): {G} $_{bare\ SiO2}$=200 and the measured Δ{G} values for the 1-L and 2-L regions shown are 15 and 33, respectively (see inset). The expected Δ{G} for 1-L and 2-L HOPG are 16 and 30, respectively (see ESM). (b) Area fraction ($\theta_{1-2LG}$) covered by no more than two graphene layers as a function of cooling rate for graphene films synthesized with Ni grain sizes $L_1$ and $L_2$. (c) Number of sites per mm$^2$ with more than two graphene layers ($\rho_{2+LG}$) vs. $dT/dt$. The two different Ni films with grain sizes $L_1$ and $L_2$ show a similar dependence on cooling rate. Optical images of the two grain sizes are shown in (d) and (e). Graphene films grown on the two types of Ni films before (f, g) and after transferring to $SiO_2$/Si (h, i). The area covered by 1–2LG is independent of Ni grain size. Scale bars in (d–i): 25 µm. (j) Photograph of a large graphene film with ~87% of its area covered by 1–2LG. The size of the films fabricated is only limited by the sizes of the Ni film and the CVD chamber employed.

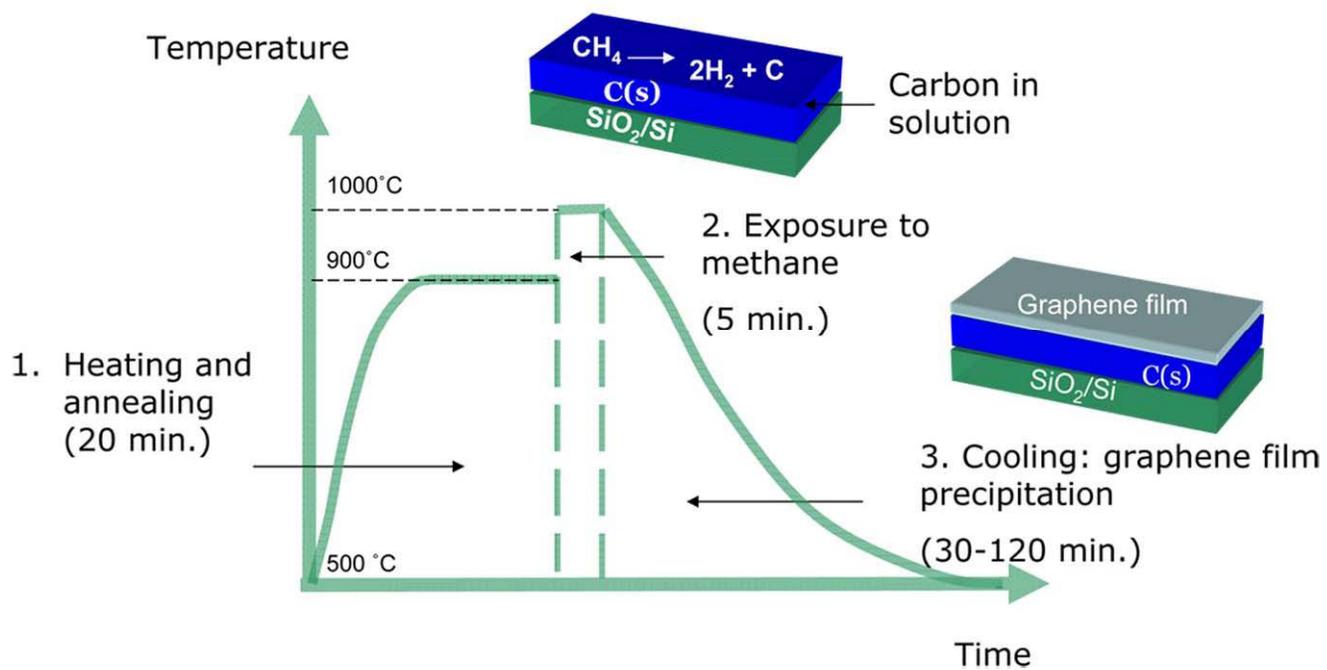

**Figure 1**

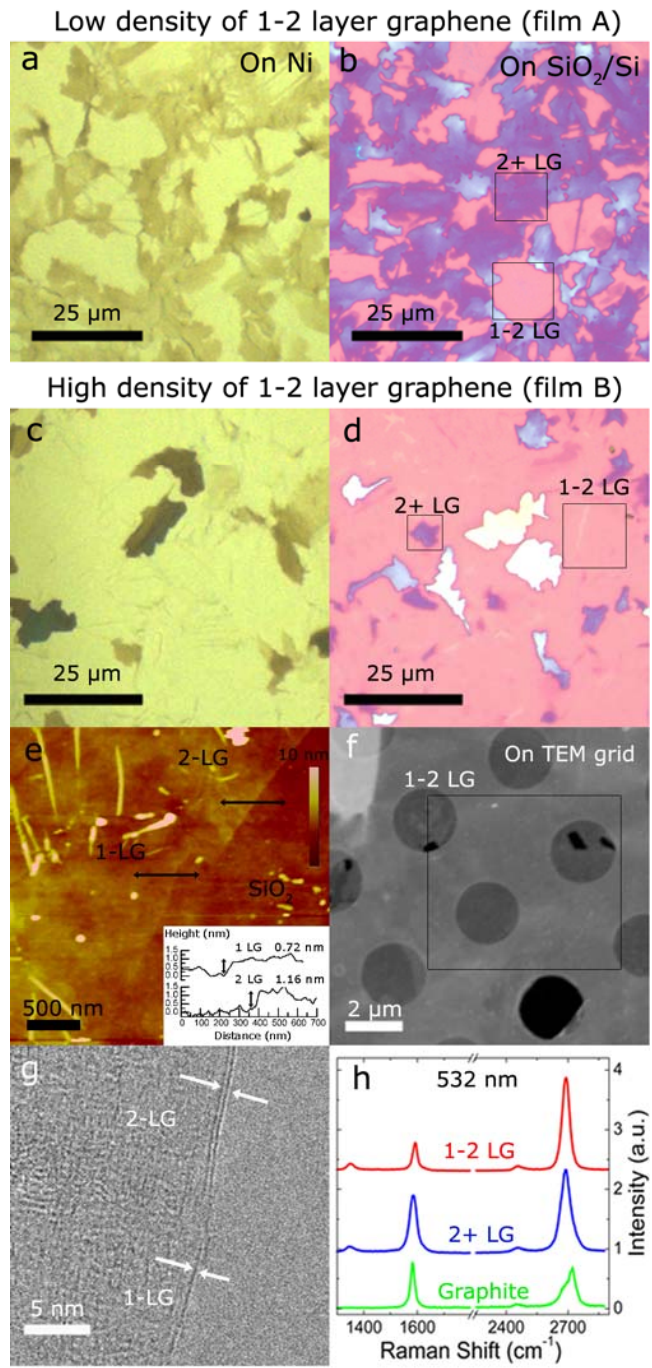

**Figure 2**

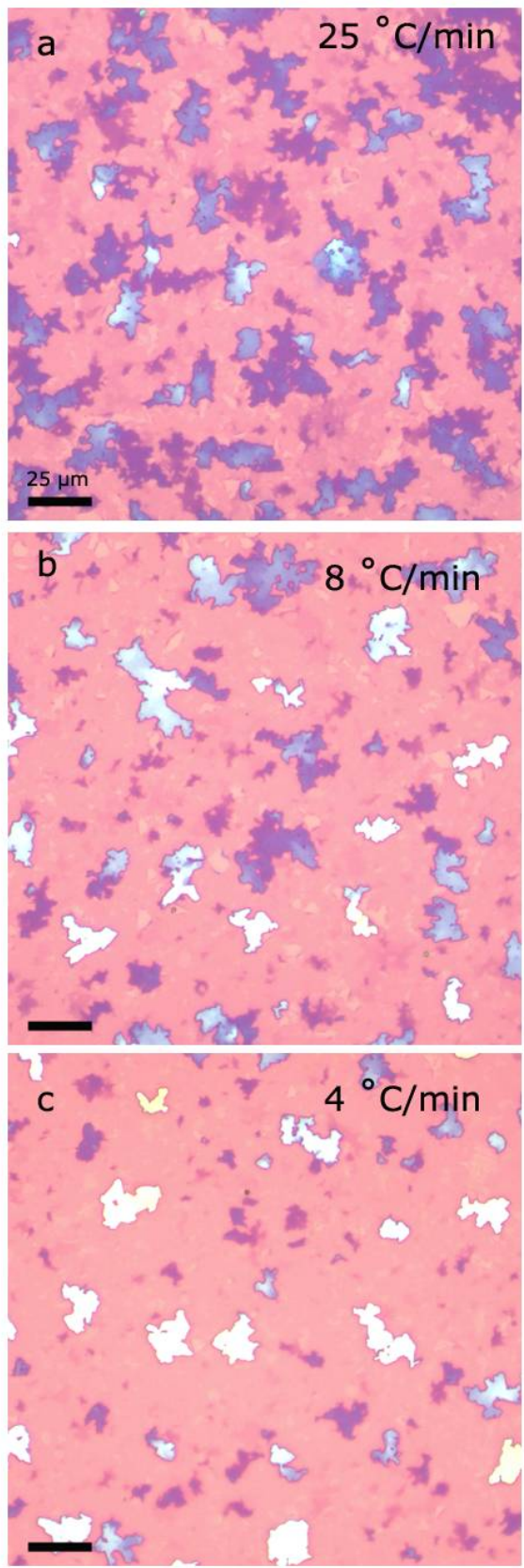

**Figure 3**

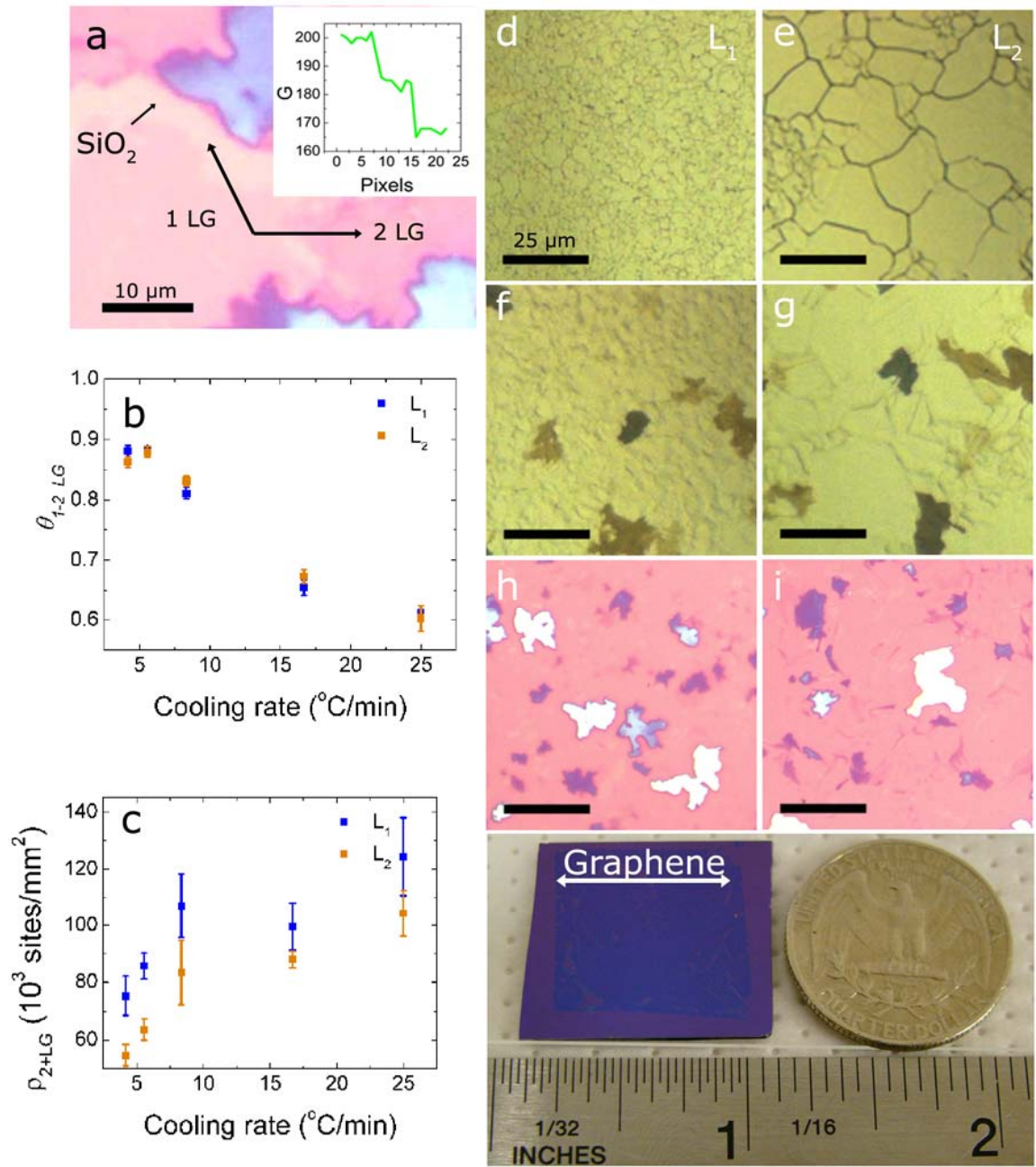

**Figure 4**

# Growth of Large-area Single- and Bi-layer Graphene by Controlled Carbon Precipitation on Polycrystalline Ni Surfaces

Electronic Supplementary Material

**I. Calibration of {G} with the number of layers using HOPG-derived graphene samples.**

We use the color contrast generated by 1-, 2- and 3-LG derived from exfoliated HOPG and deposited on 300 nm $SiO_2$/Si as an automated calibration for the assignment of the number of layers of the CVD graphene films (also on 300nm $SiO_2$/Si) as described below. Figure S-1a shows microcleaved graphene pieces on 300 nm $SiO_2$/Si. The number of layers can be determined by inspecting the G´ peak (at ~2700 $cm^{-1}$) in the Raman spectra of the graphene pieces (Fig. S-1b). AFM can also assist in the layer number assignment (Fig. S-1c). The RGB (Red Green Blue) color model is a model used for displaying and representing optical images (Foley, J. D., *Computer graphics: principles and practice.* Addison-Wesley: Reading, Mass. 1996) and is shown here to be useful for automated assignment of the layer number. In this model, each pixel of an image mixes red, green and blue light to reproduce the color of a pixel. The color obtained for each pixel depends on the intensities of the red, green and blue components that are mixed. Figure S-1d shows the values corresponding to the green component {G} extracted from the optical image along the line in Fig. S-1a. The Red, Green and Blue values of each pixel of our images are expressed on a scale of 0–255 (8-bit per channel). A stepwise change of {G} is observed (Fig. S-1d) with respect to the value of {G} corresponding to the pixels of the bare $SiO_2$/Si substrate. Each step in Fig. S-1d corresponds to the addition of one graphene layer. In Fig. S-1d, the black line shows the calculated decrease in {G} with respect to the bare $SiO_2$ {G} value. This is obtained by calculating the reflectivity of bare $SiO_2$/Si and graphene on $SiO_2$/Si at a wavelength of 532 nm (see discussion below). We define the difference between {G} of graphene layers on $SiO_2$/Si and {G} of a bare $SiO_2$/Si substrate as Δ{G} (see Fig. S-1d). We use Δ{G} to determine automatically the number of layers in each optical image as shown in Fig. S-1d.

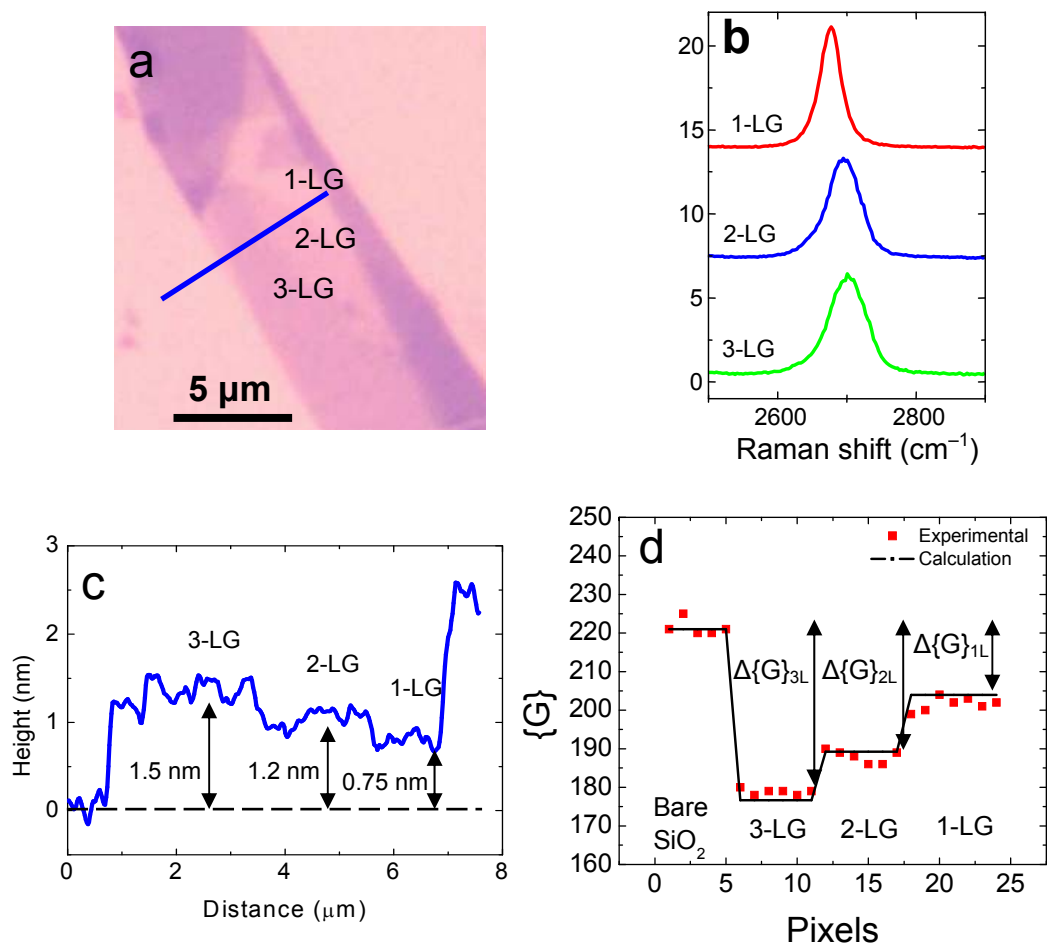

**Figure S-1** (a) An optical image of microcleaved HOPG graphene (1–3 layers) on 300 nm $SiO_2$/Si. (b) Raman spectra of the regions identified as one, two or three graphene layers in (a). (c) Measured AFM cross sectional height vs. distance corresponding to the blue line in (a). (d) {G} values extracted from the RGB values along the blue line shown in (a). The black line shows the calculated {G} values for one, two and three graphene layers on SiO$_2$ and the red points are experimental readings.

**II. Modeling {G} for graphene on SiO$_2$/Si**

To model {G} of graphene layers we consider perpendicular incident light where the magnetic field is polarized in the *z*-direction. Our system consists of three different layers, namely graphene, SiO$_2$ and Si. In each layer the magnetic field can be written as the sum of a forward and backward propagating wave (see Fig. S-2). The amplitude of the incident wave is set to be unity. The magnetic field can be expressed as (Wang, Y. Y.; Ni, Z. H.; Shen, Z. X.; Wang, H. M.; Wu, Y. H. Interference enhancement of Raman signal of graphene. *Appl. Phys. Lett.* **2008**, *92*, 043121):

$$H_{0z}(y) = e^{-ik_{0y}x} + Ae^{+ik_{0y}x} \quad (1)$$

$$H_{1z}(y) = Be^{-ik_{1y}x} + Ce^{+ik_{1y}x} \quad (2)$$

$$H_{2z}(y) = De^{-ik_{2y}x} + Fe^{+ik_{2y}x} \quad (3)$$

$$H_{3z}(y) = Ge^{-ik_{3y}x} \quad (4)$$

where *A, B, C, D, F* and *G* are parameters, $H_{iz}$ is the magnetic field in the *z*-direction and $k_{iy}$ is the wavevector in the $i^{th}$ layer, which can be calculated by:

$$k_{iy} = \frac{2\pi}{\lambda_0} n_i \quad (5)$$

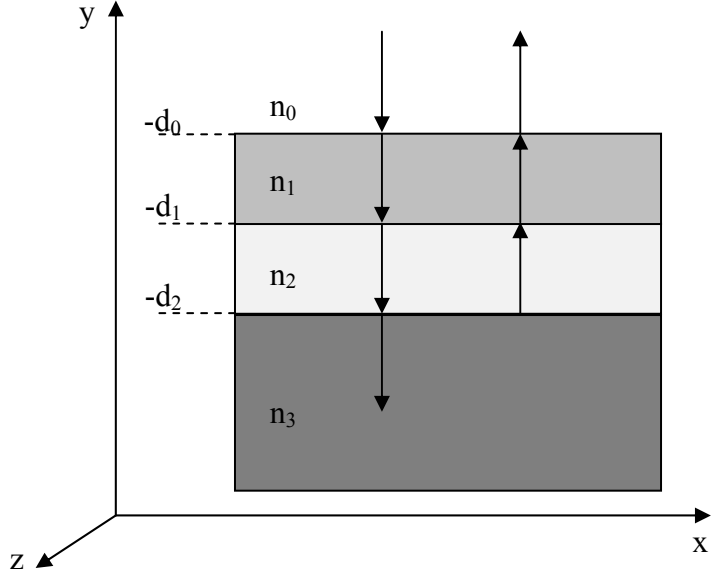

**Figure S-2** Schematic illustration of light reflection and transmission in a three layer system

where $n_i$ are the refractive indices, and $\lambda_0$ is the wavelength of the incident light (532 nm).

The boundary conditions are:

$$H_{jz}(-dj) = H_{(j+1)z}(-dj) \tag{6}$$

$$\frac{1}{\varepsilon_j}\frac{dH_{jz}}{dy}\bigg|_{-dj} = \frac{1}{\varepsilon_{j+1}}\frac{dH_{j+1z}}{dy}\bigg|_{-dj} \tag{7}$$

The thickness of graphene $d_1$ is estimated as $d_1 = m*0.335$ nm, where m is the number of layers. The thickness of $SiO_2$ is $d_2$ and the Si substrate is considered as semi-infinite: $H_{jz}$ therefore represents the magnetic field in the z-direction, - $d_j$ is the $j^{th}$ interface and $\varepsilon_j$ is the dielectric function in the $j^{th}$ layer. For this calculation the following refractive indices are used: $n_0=1$, $n_1=2.6+1.3i$, $n_2=1.46$, and $n_3=4.15+0.044i$ for air, graphite, $SiO_2$, and Si at 532 nm, respectively.

The two boundary conditions together with the three interfaces result in six equations with six variables *A, B, C, D, F* and *G* and these equations are used to determine the six variables. A is the reflectivity and is proportional to the {G} value in our optical images. To get the real {G} value of our experiment we have to multiply the reflectivity by the {G} value of the bare $SiO_2$/Si substrate, which is proportional to the illumination intensity used in the microscope settings. Now we can calculate how this value changes as a function of the number of graphene layers and the illumination conditions.

In Fig. S-3a, we plot Δ{G} as a function of the number of layers (1–3 LG) and the illumination (the {G} value of the bare $SiO_2$/Si substrate at the same microscope illumination). Points are experimental values and lines are calculated values derived from the above equations. Changes in the illumination intensities can be monitored by the {G} value of the bare $SiO_2$ background. The plot in Fig. S-3a is a plot of Δ{G} for 1-, 2- and 3-LG vs. the {G} value of the bare $SiO_2$ next to them (proportional to the illumination). The effect of illumination was considered since on the same optical image, the illumination can change depending on the pixel position (higher illumination at the center with respect to the corners). The same color scale settings are used for every image in the software utilized for acquiring them.

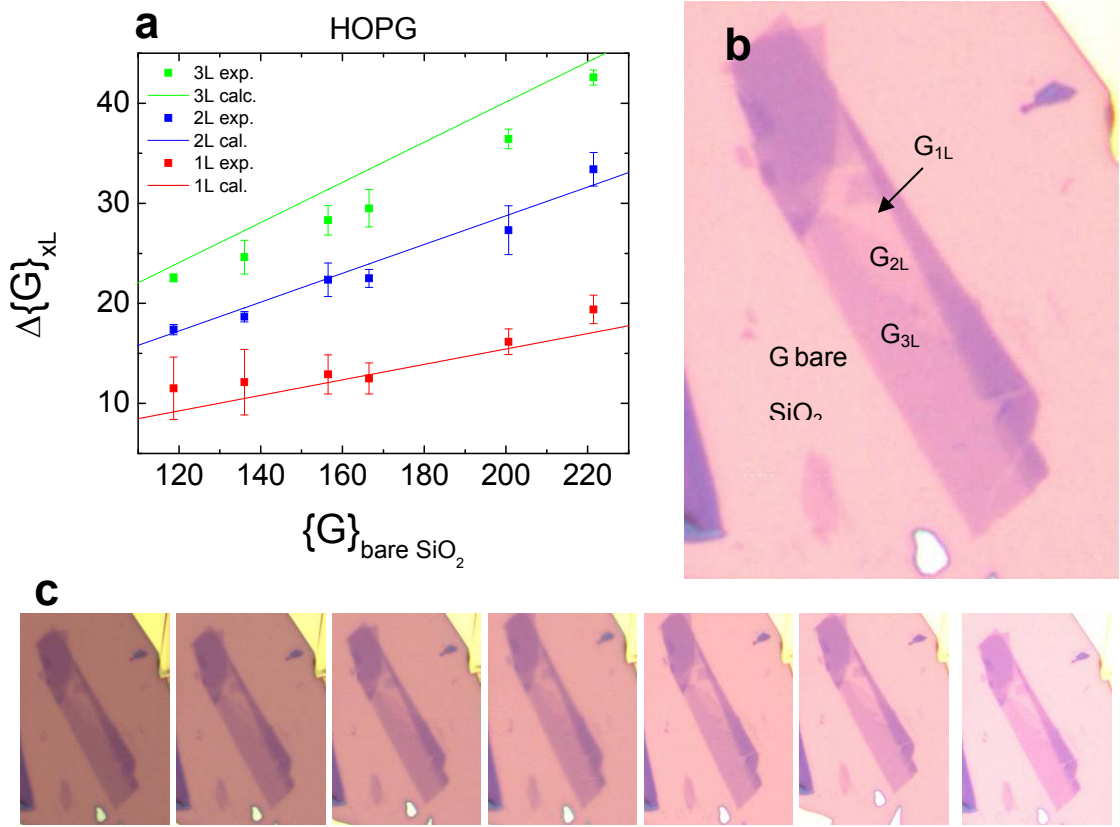

**Figure S-3** (a) Empirical dependence of $\Delta G_{xL}$ on $G_{bare\ SiO2}$ for 1-, 2- and 3-LG. Calculated results are shown as solid lines. (b) Optical image of the regions used to acquire the data for (a). (c) Examples of different illumination intensities which were used to extract the data plotted in (a).

## III. Quantification of the area covered by one and two CVD graphene layers ($\theta_{1-2LG}$) using optical images

For illustration, Fig. 4(a) in the main text and Fig. S-4 show the determination of $\Delta\{G\}$ for a specific CVD graphene region with a particular illumination. The observed values of $\Delta\{G\}$ closely match the expected values for HOPG-derived 1–2 LG at the same illumination. These assignments are also consistent with AFM height measurements.

To quantify $\theta_{1\text{-}2LG}$ for large areas, we use optical images taken at 50x magnification (field of view of 229 x 289 μm$^2$). The pixel to pixel distance is ~500 nm. The distance between thickness variations in the regions composed of 1–2 LG is usually much larger than this spatial resolution (typically on the order of a few microns). Therefore, images at this magnification and resolution are suitable for analyzing our films. Although lower magnification images could enable the quantification of $\theta_{1\text{-}2LG}$ across larger areas of the films on SiO$_2$/Si, they were not used due to the increase in pixel to pixel distance.

Identification of 1- and 2-LG was done in the following way. The {G} component of the optical images of clean SiO$_2$/Si was used as a background which is subtracted from the {G} component of the optical images of CVD graphene on SiO$_2$/Si. The background and CVD graphene images were taken at the same magnification and illumination conditions. The Δ{G} values obtained at each pixel by the subtraction were compared with the Δ{G} values expected for 1- or 2-LG (shown in Fig. S-3a) in order to label each pixel as 1- or 2-LG. This procedure was implemented with MATLAB and applied to multiple optical images in order to calculate the fraction of pixels corresponding to 1- and 2- LG in each image ($\theta_{1\text{-}2LG}$). Figure S-5 shows an example of the identification process. Notice that the pink regions of the graphene film in Fig. S-5a (1–2 LG) are identified effectively and tagged by the algorithm (Fig. S-5 b).

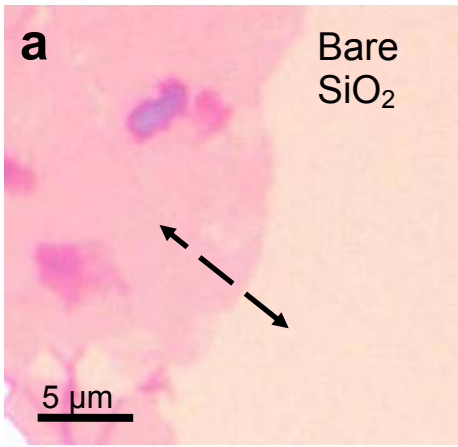
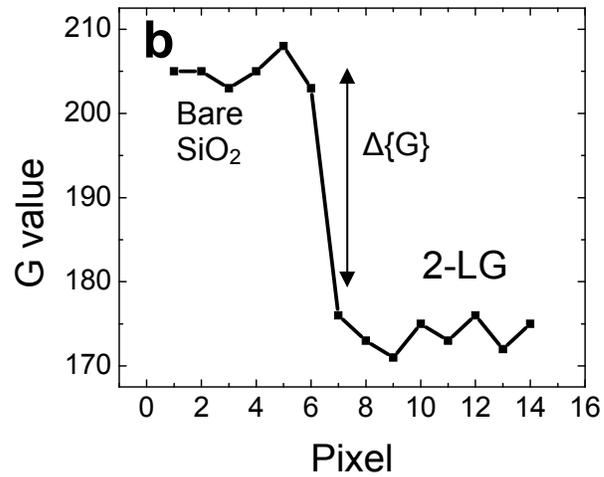

$G_{bare\ SiO_2}$=204.83 (illumination)

Δ{G}= <u>30.95</u>  (measured above)

Assignment: 2L

Expected for 2L = <u>32.23</u> (from Fig. S-3a)

**Figure S-4** (a) Optical image of another CVD graphene region on $SiO_2$/Si. (b) {G} values corresponding to the dashed line shown in (a). Δ{G} values for these regions are extracted from (b) and compared to the expected values for 1–2 LG as suggested by the fits in Fig. S-3a.

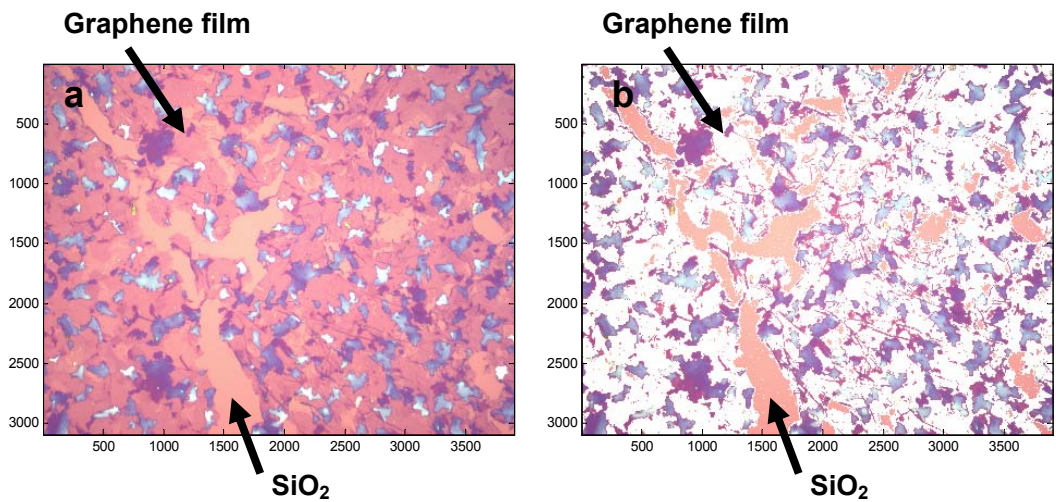

**Figure S-5** (a) An optical image of a graphene film on 300 nm $SiO_2$. (b) The same optical image as (a) with pink regions in (a) (1–2 LG) tagged with white in (b). The film was broken (shown by the upward arrow) in order to expose part of the bare $SiO_2$ substrate and to test the algorithm. Images are 290 x 230 $\mu m^2$